\documentclass{article}  
\usepackage{mycite}    
\usepackage{epsf1990}  
\twocolumn
\def\.{{\cdot}}
\def\gtapprox{\,\lower.6ex\hbox{$\buildrel >\over \sim$} \, }
\def\ltapprox{\,\lower.6ex\hbox{$\buildrel <\over \sim$} \, }

\def\arcs{\ifmmode {'' }\else $'' $\fi}     
\def\arcm{\ifmmode {' }\else $' $\fi}     
\def\deg{\ifmmode^\circ\else$^\circ$\fi}    
\def\ttimes{{\scriptstyle \times}}
\def\fr7{7$ \hskip -0.9ex \vrule height0.8ex width0.8ex depth-0.73ex
                                                                \hskip0.1ex$}

\def\hMpc{~$h^{-1}$Mpc}

\def\rinj{r_{\mbox{\small inj}}}

\newcommand\joref[5]{#1, #5, {#2, }{#3, } #4}

\newcommand\epref[3]{#1, #3, #2}

\def\MNRAS{M.N.R.A.S.}
\def\apj{Ap.J.}                 

\def\aanda{A.\&A.}            

\def\affilsize{}
\def\affilsize{\normalsize}

\title{Three-dimensional Topology-Independent Methods 
                        to Look for Global Topology}

\author{Boudewijn F. Roukema$^{1,2}$ and Vincent Blanloeil$^3$\\
{\affilsize
 $^1$Observatoire de Strasbourg, 
11, rue de l'Universit\'e, Strasbourg F-67000, France}\\
{\affilsize $^2$Nicolaus Copernicus Astronomical Center, 
ul. Bartycka 18, 00-716 Warsaw, Poland}\\
{\affilsize $^3$Institut de recherche math\'ematique avanc\'ee, 
Universit\'e Louis Pasteur et CNRS,}\\ 
{\affilsize
7 rue Ren\'e-Descartes, F-67084 Strasbourg Cedex, France}\\
{\em roukema@iap.fr, blanloei@math.u-strasbg.fr}}

\date{}

\def\Figsizes{ 
\begin{figure} 
\centerline{\epsfxsize=9cm
\epsfbox[-20 0 350 471]{"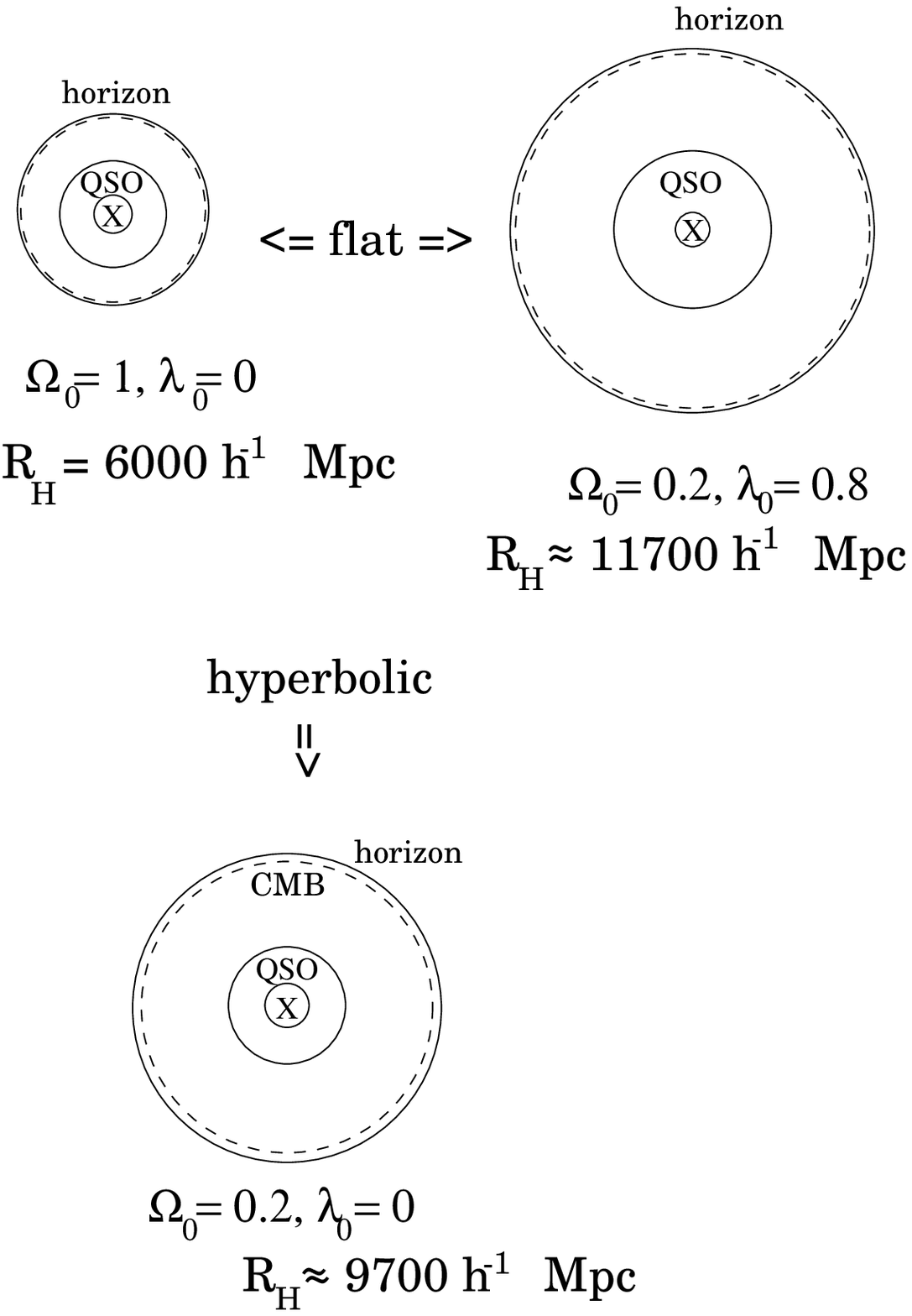"} 
}
\caption{\label{f-sizes} \protect\footnotesize
Relative sizes of observable domains of the universal covering
space, depending on
different options for the metric parameters 
$\Omega_0, \lambda_0,$ 
shown in comoving coordinates on a radially linear (proper distance) scale. 
The horizon radius $R_H$ is defined by the age of the Universe.
The inner circles for each choice of metric indicate from the
centre outwards redshifts of: $z=0\.5$ (roughly that to which we
can see X-ray emission from the richest clusters of galaxies);
$z=3$ (to which there is a high quasar density); and $z=1000$ 
(roughly the redshift of the cosmic microwave background, dashed circle).
In the flat models, the tangential distance scale is constant and the
same as the radial scales; in the hyperbolic model, the tangential 
distance scale increases (more Gpc per mm on the page) as a function
of increasing radius.
}
\end{figure} 
} 

\begin{document}

\maketitle

\begin{abstract}
The space-like hypersurface of the Universe 
at the present cosmological time is a three-dimensional 
manifold. 
A non-trivial global topology of this space-like hypersurface 
would imply that the 
apparently observable universe (the sphere of particle horizon 
radius) could contain several images of the single, physical Universe. 
Recent three-dimensional techniques for constraining and/or detecting
this topology are reviewed.
Initial applications of these techniques using X-ray bright clusters
of galaxies and quasars imply (weak) candidates 
for a non-trivial topology. 
\end{abstract}

\section{Introduction}

If the physical Universe is smaller than the ``observable Universe'',  
i.e., if the fundamental polyhedron of the Universe is smaller 
than the sphere of horizon radius in the universal covering space,
then some (or many) regions of space will be observable
at several (or many) different ``look-back'' times 
(\cite{deSitt17}~1917; \cite{Lemait58}~1958). 
The word ``Universe'' can be taken here to refer either
theoretically to the space-like hypersurface at the present
cosmological time, or observationally to the observed past time cone 
considered in comoving coordinates.
Since space-like 
hypersurfaces are three-dimensional, use of three-dimensional 
information on astrophysical objects known to exist in the covering
space provides a straight-forward way to search for or constrain
the global topology of the Universe. 

The reader is refer\-red to \cite{LaLu95} (1995) 
and to other contributions to this workshop
for an introduction to cosmological topology
and to \cite{Lum98}~(1998) for an interesting 
historical introduction. 

A brief mathematical description of the relationship between
a universal covering space $X,$ a compact 3-manifold $M$ 
and its fundamental polyhedron $P$ is provided in \S\ref{s-topol}. 
For a fuller introduction to three-dimensional geometry and
topology, see \cite{Thur82}~(1982, 1997).

Three-dimensional topology detecting techniques are based on the
required existence of multiple topological images of single physical
objects. Techniques applicable to objects observed to successively
larger scales are described in \S\ref{s-crystal} 
(``cosmic crystallography'', \cite{LLL96}~1996), 
\S\ref{s-Xclus} (``brightest X-ray clusters'', \cite{RE97}~1997) and  
\S\ref{s-isomet} (``local isometry 
detection'', \cite{Rouk96}~1996).

A Friedmann-Lema\^{\i}tre-Robertson-Walker metric (implying constant
curvature of any spatial hypersurface) is assumed 
throughout this paper. Comoving 
coordinates are used, i.e., positions of objects observed in our
past time cone are projected to the (3-D) space-like hypersurface 
at the present epoch, $t=t_0$. 
Spectroscopic redshifts, denoted $z,$ 
are used to obtain radial distance estimates 
(termed ``proper distances''
by \cite{Wein72}~1972, eq.14.2.21\footnote{The ``proper distance'' 
should not be 
confused with the quantity that
\protect\cite{Wein72}~(1972, p.485) calls ``proper motion distance'' 
and that \cite{Peeb93}~(1993, p.321, eq.13.36) 
calls ``angular size distance''.
The proper distance and proper motion distance are identical for
zero curvature, but not otherwise.}) in the $t=t_0$ 
space-like hypersurface. 

\section{Topology of Compact 3-Manifolds: Covering Spaces and 
Fundamental Polyhedra} \label{s-topol}

If we assume the Universe to have 
an FLRW metric and a trivial topology, then
it is one of the three 3-manifolds
$S^3,$ the $3$-sphere, $E^3,$ Euclidean $3$-space, or
$H^3,$ the hyperbolic $3$-space (negatively curved).
This apparent space 
is called the ``universal covering space'', which we call $X.$ 

This can be related to the real 3-manifold of the Universe, 
which we call $M$ and which can be thought of physically in terms of 
a fundamental polyhedron $P$ as follows.

One can construct a complete geometric 3-manifold by 
choosing $\Gamma$, a discrete subgroup of the group of isometries of
$X$, acting freely on $X$ (i.e., the set $\{ g\in \Gamma \ : \ gx=x \}$
is trivial for all $x$ in $X$) and take the quotient $X/\Gamma.$ 
Of course, at the present we know neither $X$ nor $\Gamma$ 
for the real Universe. 

Observations favour $X=E^3$ 
[$\Omega_0+\lambda_0=1$ in terms of standard metric parameters,
where $\lambda_0 \equiv \Lambda c^2/(3H_0^2)$] 
and $X=H^3$ ($\Omega_0+\lambda_0<1$), though as long  
as $\Omega_0+\lambda_0=1$ remains reconcilable with observations,
$X=S^3$ is likely to remain a possibility where
$\Omega_0+\lambda_0 = 1 + \epsilon$ and $\epsilon \ll 1$ 
if $\Omega_0$ and $\lambda_0$ are only measured by traditional, local
physical properties such as density. [An interesting global method
is the effect predicted by
\cite{BJS85}~(1985, p937), who
showed that 
compactness and a small amount of rotation of the observable Universe
could in principle enable the case $\Omega_0+\lambda_0 = 1$ 
to be distinguished from the case 
$\Omega_0+\lambda_0 = 1 \pm \epsilon, \,(\epsilon \ll 1)$ 
by the presence of a ``spiralling effect'' which could be seen in the
cosmological microwave background.] 

Then,
{\sl a   convex fundamental polyhedron} for the discrete
  group $\Gamma$ of the group of isometries of $X$ is a convex polyhedron $P$ 
in
   $X$ such that its interior $P^\circ$ fulfils: \nolinebreak 

i) the members of $\{ g P^\circ : g\in\Gamma\}$ are mutually disjoint,

ii) $X=\cup\{gP : g\in\Gamma\}$

iii) $\{gP : g\in\Gamma\}$ is a locally finite family of subsets of
$X$.

To find a fundamental polyhedron,
 for all $g\not=1$ in
$\Gamma$ and for $x$ in $X$, we define $H_g(x)=\{y\in X \ : \
d(x,y)<d(x,gy)\}$. 
Then {\sl the Dirichlet domain} $D(x)$, with center $x$, for 
$\Gamma$ is: $D(x)
=\cap\{H_g(x) : g\not=1 \in \Gamma\}$ when $\Gamma$ is non
  trivial and $D(x) = X$ when $\Gamma$ is trivial. The closure
  $\overline{D}(x)$ is a convex fundamental polyhedron for $\Gamma.$

Using a fundamental polyhedron $P$ in $X$, one can build a 3-manifold $M$ by 
gluing
together the sides of $P$. Poincar\'e's fundamental polyhedron theorem
proves that the inclusion of $P$ in $X$ induces an isometry from $M$
to $X/\Gamma$, where $\Gamma$ is the discrete subgroup of the group of
isometries of $X$ such that $(S,R)$ is a group presentation for
$\Gamma$ with  $S$ the set of 
sides of $P$ and
$R$ the set of relations determined by the gluing. 

When $\Gamma$ is not trivial this construction gives non-trivial
topology. 

Let us be more precise. Choose a base point $x_0$ of $X$ and
let $a : S^1 \rightarrow X/\Gamma$ be a loop based at $x_0$~;
let $b : [0,1] \rightarrow X$ be a lift of $a$ starting at $x_0$
and ending at $g_a\,x_0$ (note that $g_a$ is unique since
$\Gamma$ acts freely). The map $l : \pi_1(X/\Gamma)\rightarrow \Gamma$
defined by $l(a) = g_a$ is a homomorphism which is obviously
surjective (i.e., onto). Suppose for $a\in \pi_1(X/\Gamma)$ we have
$l(a)=1.$ Then a lift of $a$ in $\pi_1(X)$ is equal to 1
since $X$ is simply connected, hence $a=1$ and $l$ is
injective (i.e., 1:1). Therefore,  $l : \pi_1(X/\Gamma)\rightarrow
\Gamma$ is an isomorphism.

Since $\Gamma$ is a non-trivial group, the fundamental
group $\pi_1(X/\Gamma)$, which could be thought of as the group of 
non-shrinkable loops of $X/\Gamma,$ and hence of $M,$ is non-trivial. 
That is, $M$ has a non-trivial topology and can be referred to as
multi-connected.

For an example of theoretical ideas for the physical meaning of $\Gamma,$ 
see \cite{eCF98}~(1998). Here we merely consider observational 
detection of $\Gamma.$

Characterising the ``size'' of fundamental polyhedra of 3-manifolds
in a way useful observationally requires at least two parameters.
Here we adopt the ``injectivity radius'', 
$\rinj,$ i.e., half of the smallest distance from an object to 
one of its topological images; and the out-radius, $r_+,$ which is the 
radius of the smallest sphere (in the covering space) which totally
includes the fundamental polyhedron (\cite{Corn98a}~1998a). 
We refer here to $2\rinj$ as the injectivity diameter and $2r_+$ as
the out-diameter. For a discussion of these and related size parameters,  
see \cite{Corn98a}~(1998a), and note that $2\rinj$ and $2r_+$ are
similar to the parameters $\alpha$ and $\beta$ adopted by 
\cite{LaLu95}~(1995, \S10.3.3).

While both parameters represent in some sense
the ``size'' of the fundamental polyhedron, it should particularly
be remembered that many hyperbolic compact 3-D manifolds can
have $\rinj \ll r_+.$  Since we live in the plane of 
a disc galaxy --- which obscures most astronomical observations within
several degrees of the plane --- it would be difficult to
measure $2\rinj$ if it's the size of a geodesic at an angle ``close'' to
the Galactic plane.

\section{Multiple topological images of observable objects} 
\label{s-timages}

In a multi-connected universe, 
the covering space, or ``apparent'' universe, is tiled by copies 
of the fundamental polyhedron (Dirichlet domain). So the basic
principle of detecting multi-connectedness is to find multiple
``topological images''.\footnote{Also called ``topological clones''. 
The terminology ``ghosts'' is not preferred since it implies that
some images are less physically real than others.}

In a flat covering space, the particle horizon radius, $R_H,$ 
is not geometrically constrained 
to $\rinj$ and $r_+$. 
In a hyperbolic covering space,
the two quantities are both directly related to the curvature
radius, $R_C,$ (at least in the orientable cases), by 
Mostow's rigidity theorem which
states that a homotopy equivalence between two orientable hyperbolic
3-manifolds is homotopic to an isometry. 
Hence, $\rinj$ and $r_+$ are bounded below 
by $R_C$ and hence by $R_H$ to within 
a few orders of magnitude (depending on which
of the many compact hyperbolic manifolds applies to the
universe). 

So, if physical conditions in the very early Universe (quantum epoch)
tend to minimise the volume and the ``present-day'' Universe is
negatively curved, then the fundamental polyhedron may well be small
enough that multiple topological images of points of space are likely
to exist within the present-day observable sphere. Zero
curvature provides no such constraint.

Since the
Universe is of finite age, information (photons) can only be
received from within a sphere (in the covering space) 
of finite radius around the 
observer. This sphere may contain several copies of the fundamental
polyhedron, in which multiple topological images of single 
physical objects can be found.
However, objects which are seen further towards this 
``horizon'' are seen earlier in the history of the Universe, 
so are seen at different stages of the transition between a 
relatively smooth (to $\sim 0\.001\%$) material to the formation
of high-density objects such as quasars, clusters of galaxies 
and galaxies. 
So any set of observable objects can only be seen in a certain
sub-sphere (or a spherical shell) which is smaller than the
observable sphere.

Depending on the metric parameters ($\Omega_0, \lambda_0$)
of the Universe, the fraction of the comoving observable sphere
covered by a set of observable objects varies.

\Figsizes

Fig.~\ref{f-sizes} shows some characteristic distances to which
different types of objects have so far been observed in significant
quantities for a range of the metric parameters covering the
values consistent with a wide range of observational cosmological
tests. The redshifts used are indicative only.

The proper distance to a redshift $z$ can be evaluated in general as
\begin{equation}
d(z) = {c\over H_0} \int_{1/(1+z)}^{1} 
{\mbox{\rm d}a \over a \sqrt{\Omega_0/a - \kappa_0 + \lambda_0\, a^2}} 
\label{e-dprop}
\end{equation}
where $\kappa_0 \equiv \Omega_0 + \lambda_0 -1.$~\footnote{For 
readers of the popular \cite{Peeb93}~(1993), 
$\Omega_0, \lambda_0$ and $\kappa_0$ correspond to Peebles' 
$\Omega, \Omega_\Lambda$ and $-\Omega_R$ respectively.}  
This can be expressed in terms of the curvature radius 
\begin{equation}
R_C \equiv {c \over H_0 \sqrt{|\kappa_0|}} \label{e-defR_C}
\end{equation}
and the proper motion distance $d_{pm}(z)$ as
\begin{equation}
	d(z) = \left\{ 
	\begin{array}{lll}
	R_C \sinh^{-1} [d_{pm}(z)/R_C] , & \kappa_0 < 0 \\
	d_{pm}(z) , & \kappa_0 = 0 \\
	R_C \sin^{-1} [d_{pm}(z)/R_C] , & \kappa_0 > 0.
	\end{array}
	\right.
\label{e-d_dpm}
\end{equation}
If $\Omega_0 > 0$ and $\lambda_0=0,$ then the closed expression 
\begin{equation}
d_{pm}(z) = {c \over H_0} 
    { 2 [ z\Omega_0 + (\Omega_0-2)(\sqrt{\Omega_0 z + 1} -1) ]
       \over \Omega_0^{\;2} (1+z) }
\end{equation}
can be used (\cite{Wein72}~1972, p.485).
	Or, for $\kappa_0 \not=0,$ equations (\ref{e-dprop}) and 
(\ref{e-defR_C}) can be combined to give

\begin{equation}
{d(z) \over R_C} = \sqrt{|\kappa_0|} \int_{1/(1+z)}^{1} 
{\mbox{\rm d}a \over a \sqrt{\Omega_0/a - \kappa_0 + \lambda_0\, a^2}} 
\label{e-dpropR_C}
\end{equation}

Only very bright 
galaxy clusters are seen to $z=0\.5,$ and systematic ``all-sky'' surveys 
for galaxy clusters [e.g., to an X-ray luminosity of 
$L_X (0\.1-2\.4$keV$)\gtapprox 10^{45}$erg/s] are only presently being
carried out to $z \sim 0\.1.$
Quasars are in fact seen to redshifts higher than 
$z\approx 3$, but drop 
quickly in number density (e.g., \cite{Shaver98}~1998). 

The fraction of the horizon distance covered by the sphere to $z=3$ 
only decreases slightly between the models with different metric 
parameters, from 50\% in the $\Omega_0=1, \lambda_0=0$ (Einstein-de Sitter)
model to 39\% in the $\Omega_0=0\.2, \lambda_0=0$ model. If multiple
copies of the fundamental polyhedron are to be detected in a radial
direction, the relative efficiencies of 3-D methods using objects 
(e.g., quasars) visible to these redshifts does not change much.
Due to the negative curvature, however, the number of copies of 
a fundamental polyhedron which could be placed alongside one another
in a tangential direction around a sphere makes a search 
to $z\ltapprox 3$ 
much less efficient relative to a CMB 
(cosmic microwave background) search for an 
$\Omega_0=0\.2, \lambda_0=0$ universe relative to an Einstein-de Sitter
universe.

Therefore, if the Universe is as negatively curved as to give 
$\Omega_0=0\.2, \lambda_0=0,$ then 3-D searches for topological images
are going to be most efficient for multi-connected manifolds 
which have $\rinj \ll r_+.$ Since this is the case for many of the
hyperbolic compact manifolds, a 3-D method which can be applied
to objects seen to $z\ltapprox 3$ may be capable of detecting non-trivial
topology of the Universe, i.e., at least one of the generators 
$g \in \Gamma,$ in this case.

On the other hand, the flat universe model dominated by a cosmological
constant ($\Omega_0=0\.2, \lambda_0=0\.8$) shows nearly identical 
relative efficiencies of $z\ltapprox3$ and CMB methods as for the
flat, $\lambda_0=0$ model, but the relative usefulness of objects
seen to $z\ltapprox0\.5$ is much lower than in the model with a 
cosmological constant than in the model without.

\section{Previous 3-D constraints and new 3-D methods}

The out-diameter, $2r_+,$ is strongly bounded below by about
60{\hMpc} to 150{\hMpc} by the absence of 
secondary topological images of the Coma cluster of galaxies 
(\cite{Gott80}~1980) and by the existence of ``large scale structure''
(``great walls'' and filaments formed by galaxies, 
\cite{deLapp86}~1986; \cite{GH89}~1989;
\cite{daCosta93}~1993; \cite{Deng96}~1996; \cite{Einasto97}~1997).
Equivalently, this is a constraint that
$2r_+ \gtapprox R_H/100.$ 
Although mathematically not strictly excluded, it would seem 
difficult for the injectivity diameter, $2\rinj,$ to be as small as
$2\rinj \ltapprox R_H/100$ in a way that would fit the 
spatial distribution and physical properties of observed objects.

At distances greater than $R_H/10,$ the formation and evolution 
of astrophysical objects becomes much more serious than at small
scales. 

In addition, catalogues of observed objects are limited to either
wide-angle surveys to small radial distances, or 
``deep'' surveys over small solid angles. The two types of surveys 
have complementary advantages for detecting topology, though the
use of the wide-angle surveys is simpler. The increase in the characteristic
radial distances and solid angular areas of these surveys will increase
rapidly over the next few decades.

To avoid the problems of evolution, methods based primarily
on the 3-D positions of the objects (rather than their physical properties)
are needed. The methods of ``cosmic crystallography'' (\S\ref{s-crystal}) 
and of ``local isometry searching'' (\S\ref{s-isomet}) were created 
for use in catalogues of objects which are subject to evolutionary
effects on the individual astrophysical objects. If the evolutionary
effects are not too strong (and if viewing angle is not a problem),
the former method is applicable. If many objects have only a subset 
of their topological images visible due to such effects, the latter
method is necessary.

Nevertheless, the existence of a ``unique'' object (e.g., much more
brilliant than all others of its class) at a large radial distance could
still be useful in finding a lower bound to $r_+$ (particularly if 
a systematic survey over $4\pi$ steradians were available). This
idea can be used by consideration 
of the ``richest'' galaxy cluster found by
the X-ray emission emitted by its hot $T\sim 10^7K$ gas 
(\S\ref{s-Xclus}). 

For simplicity, most of the discussion below is presented in the
context of a flat, $\Omega_0=1, \lambda_0=0$ universe, but it
should be kept in mind that physical (\cite{Corn96}~1996) 
and geometrical arguments favour a hyperbolic universe.
For detailed discussions of the hyperbolic case, see the 
work of \cite{Fag85}~(1985; 1989; 1996). For numerical 
representations of compact hyperbolic manifolds 
and visualisation software, 
{\sc SnapPea} and {\sc geomview} 
(http://www.geom.umn.edu/) are recommended.

For completeness, it should also be mentioned that attempts have
been made to use the essentially two-dimensional information in the
CMB  to observationally 
bound $\rinj$ from below, by making assumptions on the 
distributions of amplitudes and phases of temperature fluctuations 
at the epoch of the CMB, by accepting foreground corrections as valid 
and by considering either particular cases
of flat geometries
(\cite{Stev93}~1993; \cite{Star93}~1993; 
\cite{JFang94}~1994; \cite{deOliv95}~1995; 
\cite{Levin98}~1998)
or individual cases of hyperbolic geometries 
(\cite{BPS98}~1998). If the caveats of
these techniques are accepted as correct, then 
$\rinj$ seems not much smaller than the 
horizon size. Indeed, the latter authors find a candidate 
hyperbolic manifold, for $\Omega_0=0\.8,$ which is
``preferable to standard CDM'' relative to the observed CMB
(\cite{BPS98}, \S4.3), but the volume of its fundamental polyhedron 
is slightly larger than that of the observable sphere. 

It should also be noted that non-trivial topology is usually 
adopted in $N$-body simulations of the formation of galaxies and
large-scale structure, but for numerical rather than physical
reasons (see \cite{FMel}~1990 for an explicit analysis).

\subsection{Cosmic crystallography} \label{s-crystal}

In a catalogue of objects in which multiple topological images 
of single physical objects are often seen, a histogram of 
object-object pair separations [for all $N(N+1)/2$ pairs in a set
of $N$ objects] should show sharp peaks due to pairs of topological
images separated by multiples of the vectors which generate 
the fundamental polyhedron from the covering space. 
\cite{LLL96}~(1996) used simulations to show that this method
should be efficient and independent of topology, at least for 
the case of zero curvature. The application of this method to 
the classical Abell and ACO cluster catalogues, to $z\approx 0\.25,$ 
didn't show any obvious topological signal (but see also 
\cite{FG97}~1997).
The method was devised to also work in the cases of non-zero
curvature, but its practical application under astronomical
conditions has yet to be carried out.

The Abell and ACO catalogues will soon be superceded by X-ray
selected ``all-sky''\footnote{``All-sky'' 
can mean as little as 2/3, though
usually more, of $4\pi$ steradians, due to obscuration by the Galaxy.}
catalogues which suffer less serious systematic biases,
but only to $z \sim 0\.1$ in programs already in progress 
for finding rich clusters.

The intrinsically brightest of these clusters
should be found to a factor of several higher in redshift, though
in numbers too small for the histogram method to show any peaks. 
In this case, the following method can be applied.

\subsection{X-ray clusters as ``standard candles''} \label{s-Xclus}

If a small number of the brightest objects of a given class 
(in particular, galaxy clusters selected in X-rays) are known
out to a given redshift, these can be considered to be unique
objects if their evolutionary properties are simple enough.

In the case of the richest galaxy clusters, which are dominated
by hot gas, these objects are unlikely to become any less luminous
as time increases, though they are likely to increase somewhat 
in luminosity. This is because galaxy clusters are the largest 
gravitationally bound objects which have had time to collapse
in the age of the Universe, and are dominated by their hot 
hydrogen gas which is in approximate kinetic equilibrium. 
Within cosmologically available time scales, it is difficult to
see how a high enough fraction of 
this gas could either cool, escape or turn into galaxies
in order for a secondary topological image at a lower redshift 
(i.e., at a more recent epoch) to be invisible in X-rays.

\mbox{\cite{RE97}}~(1997) noticed that 
\hbox{RX~J1347.5}\hbox{-1147}, which is probably 
the brightest X-ray cluster known 
(in the $0\.1-2\.4$keV frequency band) is quite distant from us 
(with respect to other known galaxy clusters). So, if we could 
be sure that there were no
topological images of this cluster closer to us in any direction, 
then the distance 
to this cluster would give the lower limit 
$r_+ \gtapprox 1100\pm100${\hMpc}.  
(The uncertainty is due to the observational uncertainty in 
the metric parameters; the range shown in Fig.~\ref{f-sizes} 
is adopted.) However, since galactic obscuration is important, 
this is strictly speaking a weaker limit, i.e., 
$2r_+ \gtapprox 1100\pm100${\hMpc}.  

Because the object is unique,
the only topological image pair geodesics which are excluded 
are those extending from the known image 
to the borders of the observed volume. Small closed geodesics in other
directions, e.g., roughly perpendicular to any large 
geodesic running from 
RX~J1347.5-1147 to a distant point in the
north galactic cone defined by $b^{II} > 20\deg, z \le 0\.451,$ 
would not contradict the existence of RX~J1347.5-1147 as an 
(apparently) unique object. Hence, this method only constrains 
$2r_+$ rather than $2\rinj.$

Serendipitously, a candidate for the topology was noticed
by comparing the 3-D positions of the several bright clusters
listed by \cite{RE97}~(1997). Three (of the seven clusters studied) 
form a right angle (to 2\% accuracy) with side lengths equal within
1\%. This is just what would be expected in the case of 
a $T^2 \ttimes X$ manifold where $X$ is unknown. There is no obvious
physical motivation for this case to be favoured, though it is 
commonly used for pedagogical purposes, 
as in several of the early
attempts at trying to constrain topology using the CMB as observed
by the COBE satellite. 

\mbox{\protect\cite{RE97}}~(1997) list several arguments against 
topological identity of the three clusters, 
but the cleanest observational test
would be to verify or refute the existence of further implied 
topological images. While further implied topological images
could exist at high redshifts, the cluster might not have formed
at those early epochs. So implied images at low 
enough redshifts that the cluster is guaranteed to be in existence 
with a minimum luminosity should be considered.

Specific predictions of this (weak) candidate 3-manifold containing 
two known generators $g_1, g_2$ and one unknown generator 
$g_3$ are as follows.
If this candidate as suggested by 
Roukema \& Edge
(1997) were
correct, then the object seen by ROSAT, 
RX J203150.4 -403656, should be a galaxy cluster at $0\.38$ $< z$ $< 0\.40,$
and the Arp-Madore galaxy cluster  AM 0750 -490 (which would be the 
low redshift topological image of 
 MS 1054 -0321) would be at $0\.23$ $< z$ $< 0\.26.$ 

\subsection{Local isometry searches} \label{s-isomet}

The previous methods can only be applied to objects whose evolution
is relatively simple. The objects which can be easily seen to 
$z\sim3$ in significant numbers, quasars, have evolutionary 
properties which are not well understood. They are likely to 
have short lifetimes, either with recurrent bursts related 
to (maybe) mergers of galaxy dark matter haloes or differing 
lifetimes depending on individual quasar properties. In addition,
according to the ``unified model'' of active galactic nuclei (AGN),
a quasar seen from a very different angle appears much fainter,
as a Seyfert galaxy, for example, with a very small chance of having
already been observed in high-redshift galaxy searches.

One approach to using quasars is to consider special cases. 
\cite{Fag87}~(1987) searched for images of the Galaxy seen as
quasars in directions separated by 
180\deg or 90\deg.

A more general method is to accept the problem of multiple topological 
image visibility and
search in a large catalogue for the rare cases in which {\em several}
objects in two different topological images of a single physical 
3-D region are both visible. In other words, one searches 
for an isometry between 
two regions of the covering space, each of a few hundred {\hMpc} 
in size. 

If several such isometries are found, then these should 
be used to generate the full set of transformations from the
covering space to the fundamental polyhedron. These would then
be confirmed (or refuted) by the predictions of the 3-D positions
of multiple topological images of other quasars, or by comparison
with the CMB.

This method was first presented by 
\protect\cite{Rouk96}~(1996) and applied to a  
catalogue of $N\approx $~5000 
quasars at $z>1.$ Two isometries, i.e., 
two pairs of quasar quintuplets separated by more than 
300\hMpc, were found. Due to the number density distribution 
of quasars, this is not necessarily due to topological imaging.
Simulated catalogues showed that there is about a 30\% chance
of finding two similar coincidences in a universe of trivial 
topology with the same observational selection criteria for
finding quasars.

At best, these two isometries could be considered as defining 
a weak candidate for the 3-manifold in which we live. 
This candidate would be non-orientable. 

The technique as presented by \cite{Rouk96}~(1996) was 
only an initial implementation of the basic principle. 
The parameters chosen may not necessarily be optimal 
for obtaining a detection. 

For example, although the number of isometries of 
$n$-tuplets is higher when $n$ is lower, 
it may well be possible that the signal (due to topological isometries)
may increase faster than the noise (due to the number density distribution
of the catalogue for a simply connected universe)
when $n$ is lowered, so that the signal-to-noise ratio 
would increase.

Another use of the technique using the same size data set 
would be to test a series of universe 
models with incrementally different metric parameters
(e.g., $\Omega_0=0\.20, 0\.21, 0\.22, ..., 1\.10$),
in particular the negatively curved models. The increment 
should be chosen as a function of the uncertainty regarding 
quasars' (3-D) positions.

Alternatively, one could allow for a scaling factor when 
comparing $n$-tuplets, in which case isometries due to any 
value of the curvature would automatically be detected. 
However, this would also increase the number of chance coincidences
of $n$-tuplets, and is not strictly correct --- it 
would distort the local geometrical relations 
if the $n$-tuplets are not small enough.

In the method of adopting successive values of 
the metric parameters, the successive signal-to-noise ratios 
(numbers of isometric $n$-tuplet pairs due to topology compared to 
numbers due to chance)
are $S_i/N_i$ where 
\begin{equation}
S_i  \left\{
\begin{array}{lll}
 = 0 , & (i \not\approx i^*)\\
 > 0 , & (i \approx i^*)
\end{array}
\right.
\end{equation} 
and $(\Omega_0, \lambda_0)_{i^*}$ are the metric parameters of the
real Universe. The signal-to-noise ratio $S_i/N_i$ 
is then zero except when the metric parameters are close to correct,
i.e., $i\approx i^*,$ and has a maximum of 
$S_{i^*} / N_{i^*}.$ (The values of the
metric parameters would then be quite tightly constrained.)

If a scaling factor is used instead, then the calculation
should be much faster (since a single set of metric parameters 
is adopted), but the signal-to-noise ratio 
becomes
\begin{equation}
\begin{array}{lll}
{\sum_i S_i \over
 \sum_i N_i + \sum_{i\not= j} N_{ij}} &= & 
{\sum_{i\approx i^*} S_i \over \sum_i N_i + \sum_{i\not= j} N_{ij}} \\
& \approx& {S_{i^*} \over \sum_i N_i + \sum_{i\not= j} N_{ij} } \\
& \ll&  S_{i^*} / N_{i^*},
\end{array}
\end{equation}
where $N_{ij}$ are numbers of chance coincidences for scaling factors
which imply inconsistent curvature estimates.
So, 
the signal-to-noise ratio is much lower if scaling is used. 
Even if the scaling is
done requiring consistent values of the curvature between $n$-tuplets
at different redshifts (which would probably slow down the calculation),
the signal-to-noise ratio would still be lower than for the successive
metric value method. Additionally, distortions due to the greater than
zero size of $n$-tuplets would make $S_{i^*}$ in the scaling method
slightly lower than for the successive metric method.

The other development of this method is that 
since quasars are visible to about $R_H/2,$ 
future quasar 
surveys will provide a more thorough sampling of this volume,
so that the ``rare'' cases searched for will increase in number
and the chances of detection (if the topology is detectable) 
will increase.

\section{Conclusion}

Several methods have been developed in the last few years to
either detect or constrain the topology of the 
spatial part of the Universe. The
relative efficiency (in terms of fundamental polyhedron crossings) 
of 3-D to 2-D methods depends moderately 
on the precise values of the metric parameters $\Omega_0, \lambda_0.$ 
Objects seen to about $z\sim3$ would cross half the horizon
distance for any presently accepted metric parameters, 
and in a cosmological constant dominated universe, 
objects seen to $z\sim 0\.1 -0\.5$ would cover many fewer copies of
the fundamental polyhedron than the CMB.

Initial applications of 3-D methods to existing observational 
catalogues or individual observations indicate several (weak) candidates
for the 3-manifold in which we live. (Or more precisely, for 
some of the generators of the 3-manifold.)
These candidates are falsifiable
with moderate observational investment in telescope time.

Moreover, further development of the local isometry search method 
is presently possible for application to existing observational 
quasar catalogues.

Looking to the future, new catalogues of objects
made over the next few years, 
in particular all-sky surveys of quasars, 
will possibly allow the topology of the 
Universe to be detected to a high significance by the
local isometry search method. Alternatively, the ``circles method'' 
of \cite{Corn98b}~(1998b) applied to the observations by 
either MAP or Planck (planned CMB satellites) is likely to either reveal or
constrain the topology of the Universe.

Within a decade, we should know whether or not the topology of the 
Universe is detectable, and if so what it is.

This research has been partially supported by the 
Polish Council for Scientific Research Grant
KBN 2 P03D 008 13.


\end{document}